\documentclass[
    aps,
    prd,
    showpacs,
    letterpaper,
    reprint,
    notitlepage,
    nofootinbib,
    floatfix,
    amsmath,
    superscriptaddress
]{revtex4-2}

\usepackage[utf8]{inputenc}
\usepackage[ngerman,english]{babel}

\usepackage[colorlinks=true,allcolors=blue]{hyperref}

\usepackage{physics}
\usepackage{graphicx}
\usepackage{cleveref}
\usepackage{siunitx}
\usepackage{booktabs}

\usepackage{pgfplots}
\usepackage{pgfplotstable}
\ifluatex
\else
    \DeclareUnicodeCharacter{2212}{−}
\fi
\usepgfplotslibrary{groupplots,dateplot}
\usetikzlibrary{patterns,shapes.arrows, decorations.markings,shapes.geometric}
\pgfplotsset{compat=newest}
\usepgfplotslibrary{fillbetween}

\definecolor{black}{RGB}{0,0,0}
\definecolor{orange}{RGB}{230, 159, 0}
\definecolor{skyblue}{RGB}{86, 180, 233}
\definecolor{bluishgreen}{RGB}{0, 158, 115}
\definecolor{yellow}{RGB}{240, 228, 66}
\definecolor{blue}{RGB}{0, 114, 178}
\definecolor{vermilion}{RGB}{213, 94, 0}
\definecolor{reddishpurple}{RGB}{204, 121, 167}

\definecolor{lightgray204}{RGB}{204,204,204}

\usepackage{helvet}
\usepackage[eulergreek]{sansmath}
\pgfplotsset{
    tick label style = {font=\footnotesize\sansmath\sffamily},
    every axis label = {font=\footnotesize\sansmath\sffamily},
    label style = {font=\footnotesize\sffamily},
    legend style = {
        font=\footnotesize\sffamily,
        fill opacity=0.8,
        draw opacity=1,
        text opacity=1,
        at={(0.5,0.95)},
        anchor=north,
        draw=lightgray204,
        /tikz/every even column/.append style={column sep=0.1cm}
    },
    legend cell align={left}
}
\tikzstyle{plot annotation}=[
    font=\footnotesize\sffamily,
    fill opacity=0.8,
    draw opacity=1,
    text opacity=1,
    draw=lightgray204,
    fill=white
]

\usepgfplotslibrary{external}
\tikzset{external/only named=true}

\begin{document}

%TC:ignore
\title{Search for a light sterile neutrino with 7.5 years of IceCube DeepCore data}

% -------- AUTHORS ---------
\affiliation{III. Physikalisches Institut, RWTH Aachen University, D-52056 Aachen, Germany}
\affiliation{Department of Physics, University of Adelaide, Adelaide, 5005, Australia}
\affiliation{Dept. of Physics and Astronomy, University of Alaska Anchorage, 3211 Providence Dr., Anchorage, AK 99508, USA}
\affiliation{Dept. of Physics, University of Texas at Arlington, 502 Yates St., Science Hall Rm 108, Box 19059, Arlington, TX 76019, USA}
\affiliation{CTSPS, Clark-Atlanta University, Atlanta, GA 30314, USA}
\affiliation{School of Physics and Center for Relativistic Astrophysics, Georgia Institute of Technology, Atlanta, GA 30332, USA}
\affiliation{Dept. of Physics, Southern University, Baton Rouge, LA 70813, USA}
\affiliation{Dept. of Physics, University of California, Berkeley, CA 94720, USA}
\affiliation{Lawrence Berkeley National Laboratory, Berkeley, CA 94720, USA}
\affiliation{Institut f{\"u}r Physik, Humboldt-Universit{\"a}t zu Berlin, D-12489 Berlin, Germany}
\affiliation{Fakult{\"a}t f{\"u}r Physik {\&} Astronomie, Ruhr-Universit{\"a}t Bochum, D-44780 Bochum, Germany}
\affiliation{Universit{\'e} Libre de Bruxelles, Science Faculty CP230, B-1050 Brussels, Belgium}
\affiliation{Vrije Universiteit Brussel (VUB), Dienst ELEM, B-1050 Brussels, Belgium}
\affiliation{Department of Physics and Laboratory for Particle Physics and Cosmology, Harvard University, Cambridge, MA 02138, USA}
\affiliation{Dept. of Physics, Massachusetts Institute of Technology, Cambridge, MA 02139, USA}
\affiliation{Dept. of Physics and The International Center for Hadron Astrophysics, Chiba University, Chiba 263-8522, Japan}
\affiliation{Department of Physics, Loyola University Chicago, Chicago, IL 60660, USA}
\affiliation{Dept. of Physics and Astronomy, University of Canterbury, Private Bag 4800, Christchurch, New Zealand}
\affiliation{Dept. of Physics, University of Maryland, College Park, MD 20742, USA}
\affiliation{Dept. of Astronomy, Ohio State University, Columbus, OH 43210, USA}
\affiliation{Dept. of Physics and Center for Cosmology and Astro-Particle Physics, Ohio State University, Columbus, OH 43210, USA}
\affiliation{Niels Bohr Institute, University of Copenhagen, DK-2100 Copenhagen, Denmark}
\affiliation{Dept. of Physics, TU Dortmund University, D-44221 Dortmund, Germany}
\affiliation{Dept. of Physics and Astronomy, Michigan State University, East Lansing, MI 48824, USA}
\affiliation{Dept. of Physics, University of Alberta, Edmonton, Alberta, T6G 2E1, Canada}
\affiliation{Erlangen Centre for Astroparticle Physics, Friedrich-Alexander-Universit{\"a}t Erlangen-N{\"u}rnberg, D-91058 Erlangen, Germany}
\affiliation{Physik-department, Technische Universit{\"a}t M{\"u}nchen, D-85748 Garching, Germany}
\affiliation{D{\'e}partement de physique nucl{\'e}aire et corpusculaire, Universit{\'e} de Gen{\`e}ve, CH-1211 Gen{\`e}ve, Switzerland}
\affiliation{Dept. of Physics and Astronomy, University of Gent, B-9000 Gent, Belgium}
\affiliation{Dept. of Physics and Astronomy, University of California, Irvine, CA 92697, USA}
\affiliation{Karlsruhe Institute of Technology, Institute for Astroparticle Physics, D-76021 Karlsruhe, Germany}
\affiliation{Karlsruhe Institute of Technology, Institute of Experimental Particle Physics, D-76021 Karlsruhe, Germany}
\affiliation{Dept. of Physics, Engineering Physics, and Astronomy, Queen's University, Kingston, ON K7L 3N6, Canada}
\affiliation{Department of Physics {\&} Astronomy, University of Nevada, Las Vegas, NV 89154, USA}
\affiliation{Nevada Center for Astrophysics, University of Nevada, Las Vegas, NV 89154, USA}
\affiliation{Dept. of Physics and Astronomy, University of Kansas, Lawrence, KS 66045, USA}
\affiliation{Centre for Cosmology, Particle Physics and Phenomenology - CP3, Universit{\'e} catholique de Louvain, Louvain-la-Neuve, Belgium}
\affiliation{Department of Physics, Mercer University, Macon, GA 31207-0001, USA}
\affiliation{Dept. of Astronomy, University of Wisconsin{\textemdash}Madison, Madison, WI 53706, USA}
\affiliation{Dept. of Physics and Wisconsin IceCube Particle Astrophysics Center, University of Wisconsin{\textemdash}Madison, Madison, WI 53706, USA}
\affiliation{Institute of Physics, University of Mainz, Staudinger Weg 7, D-55099 Mainz, Germany}
\affiliation{Department of Physics, Marquette University, Milwaukee, WI 53201, USA}
\affiliation{Institut f{\"u}r Kernphysik, Westf{\"a}lische Wilhelms-Universit{\"a}t M{\"u}nster, D-48149 M{\"u}nster, Germany}
\affiliation{Bartol Research Institute and Dept. of Physics and Astronomy, University of Delaware, Newark, DE 19716, USA}
\affiliation{Dept. of Physics, Yale University, New Haven, CT 06520, USA}
\affiliation{Columbia Astrophysics and Nevis Laboratories, Columbia University, New York, NY 10027, USA}
\affiliation{Dept. of Physics, University of Oxford, Parks Road, Oxford OX1 3PU, United Kingdom}
\affiliation{Dipartimento di Fisica e Astronomia Galileo Galilei, Universit{\`a} Degli Studi di Padova, I-35122 Padova PD, Italy}
\affiliation{Dept. of Physics, Drexel University, 3141 Chestnut Street, Philadelphia, PA 19104, USA}
\affiliation{Physics Department, South Dakota School of Mines and Technology, Rapid City, SD 57701, USA}
\affiliation{Dept. of Physics, University of Wisconsin, River Falls, WI 54022, USA}
\affiliation{Dept. of Physics and Astronomy, University of Rochester, Rochester, NY 14627, USA}
\affiliation{Department of Physics and Astronomy, University of Utah, Salt Lake City, UT 84112, USA}
\affiliation{Dept. of Physics, Chung-Ang University, Seoul 06974, Republic of Korea}
\affiliation{Oskar Klein Centre and Dept. of Physics, Stockholm University, SE-10691 Stockholm, Sweden}
\affiliation{Dept. of Physics and Astronomy, Stony Brook University, Stony Brook, NY 11794-3800, USA}
\affiliation{Dept. of Physics, Sungkyunkwan University, Suwon 16419, Republic of Korea}
\affiliation{Institute of Basic Science, Sungkyunkwan University, Suwon 16419, Republic of Korea}
\affiliation{Institute of Physics, Academia Sinica, Taipei, 11529, Taiwan}
\affiliation{Dept. of Physics and Astronomy, University of Alabama, Tuscaloosa, AL 35487, USA}
\affiliation{Dept. of Astronomy and Astrophysics, Pennsylvania State University, University Park, PA 16802, USA}
\affiliation{Dept. of Physics, Pennsylvania State University, University Park, PA 16802, USA}
\affiliation{Dept. of Physics and Astronomy, Uppsala University, Box 516, SE-75120 Uppsala, Sweden}
\affiliation{Dept. of Physics, University of Wuppertal, D-42119 Wuppertal, Germany}
\affiliation{Deutsches Elektronen-Synchrotron DESY, Platanenallee 6, D-15738 Zeuthen, Germany}

\author{R. Abbasi}
\affiliation{Department of Physics, Loyola University Chicago, Chicago, IL 60660, USA}
\author{M. Ackermann}
\affiliation{Deutsches Elektronen-Synchrotron DESY, Platanenallee 6, D-15738 Zeuthen, Germany}
\author{J. Adams}
\affiliation{Dept. of Physics and Astronomy, University of Canterbury, Private Bag 4800, Christchurch, New Zealand}
\author{S. K. Agarwalla}
\thanks{also at Institute of Physics, Sachivalaya Marg, Sainik School Post, Bhubaneswar 751005, India}
\affiliation{Dept. of Physics and Wisconsin IceCube Particle Astrophysics Center, University of Wisconsin{\textemdash}Madison, Madison, WI 53706, USA}
\author{J. A. Aguilar}
\affiliation{Universit{\'e} Libre de Bruxelles, Science Faculty CP230, B-1050 Brussels, Belgium}
\author{M. Ahlers}
\affiliation{Niels Bohr Institute, University of Copenhagen, DK-2100 Copenhagen, Denmark}
\author{J.M. Alameddine}
\affiliation{Dept. of Physics, TU Dortmund University, D-44221 Dortmund, Germany}
\author{N. M. Amin}
\affiliation{Bartol Research Institute and Dept. of Physics and Astronomy, University of Delaware, Newark, DE 19716, USA}
\author{K. Andeen}
\affiliation{Department of Physics, Marquette University, Milwaukee, WI 53201, USA}
\author{C. Arg{\"u}elles}
\affiliation{Department of Physics and Laboratory for Particle Physics and Cosmology, Harvard University, Cambridge, MA 02138, USA}
% ...
\date{\today}

\collaboration{IceCube Collaboration}
\noaffiliation

%%%%% ABSTRACT %%%%%%%
\begin{abstract}
    We present a search for an eV-scale sterile neutrino using 7.5 years of data from the IceCube DeepCore detector. The analysis uses a sample of 21,914 events with energies between 5 and 150 GeV to search for sterile neutrinos through atmospheric muon neutrino disappearance. Improvements in event selection and treatment of systematic uncertainties provide greater statistical power compared to previous DeepCore sterile neutrino searches. Our results are compatible with the absence of mixing between active and sterile neutrino states, and we place constraints on the mixing matrix elements $|U_{\mu 4}|^2 < 0.0534$ and $|U_{\tau 4}|^2 < 0.0574$ at 90\% CL under the assumption that $\Delta m^2_{41}\geq \SI{1}{eV^2}$. These null results add to the growing tension between anomalous appearance results and constraints from disappearance searches in the 3+1 sterile neutrino landscape.
\end{abstract}

\maketitle
%TC:endignore

%%%%%%% Introduction %%%%%%%%%%

\section{Introduction}

While the three-neutrino oscillation framework has been remarkably successful in explaining most observations of neutrino flavor transitions, several anomalies have emerged from various experiments that cannot be reconciled within this paradigm. Notably, the LSND and MiniBooNE experiments have reported an excess of electron-neutrino-like events in muon neutrino beams \cite{LSND:2001aii,miniboone2021}, which could be explained by the existence of a fourth, sterile neutrino species with a mass splitting of approximately \SI{1}{eV^2} relative to the three active neutrino flavors. This interpretation is further supported by the long-standing Gallium anomaly \cite{gallex:1998}, a deficit of electron-neutrinos observed in radioactive source experiments that has been recently confirmed by the BEST and SAGE experiments \cite{Barinov_2022,Abdurashitov_2006}. These intriguing anomalies have motivated extensive efforts to search for sterile neutrinos in the eV mass range.

Although the sterile neutrinos do not directly interact via the weak force, they can mix with  the active neutrino flavor eigenstates in a way that influences their oscillation behavior. The simplest mathematical description of the effect is the so-called 3+1 model, in which a fourth mass eigenstate with mass splitting $\Delta m^2_{41}$ and a non-interacting flavour eigenstate $\nu_s$ is added to the standard three-flavour model. The PMNS matrix \cite{pmns_matrix} is extended by a fourth row and column, such that 
\begin{equation}
U_{\text{PMNS}}^{3+1} = \begin{pmatrix}
U_{e1} & U_{e2} & U_{e3} & U_{e4} \\
U_{\mu 1} & U_{\mu 2} & U_{\mu 3} & U_{\mu 4} \\
U_{\tau 1} & U_{\tau 2} & U_{\tau 3} & U_{\tau 4} \\
U_{s1} & U_{s2} & U_{s3} & U_{s4}
\end{pmatrix}\;.
\end{equation}
The matrix elements $U_{\ell 4}$ determine the amount of mixing between the neutrino flavor $\ell$ and the fourth mass eigenstate.
In this paradigm, the $\nu_\mu \rightarrow \nu_e$ flavor transition probability for short baseline experiments such as LSND and MiniBooNE is approximately proportional to the product $|U_{\mu 4}|^2|U_{e4}|^2$, while the $\nu_e$ survival probability measured in Gallium experiments depends only on the value of $|U_{e4}|^2$ \cite{Dentler_2018}.

The landscape of experimental tests of this model currently shows a highly conflicted picture. While the aforementioned anomalies are statistically highly significant, the mixing amplitudes $|U_{e4}|$ and $|U_{\mu4}|$ that would be necessary to explain them are in strong tension with the combined non-anomalous measurements of the disappearance channels $\nu_\mu\rightarrow\nu_\mu$ and $\nu_e \rightarrow \nu_e$, which favor standard three-flavor neutrino mixing \cite{Dentler_2018,Diaz_2020}.  This includes previous measurements performed by IceCube DeepCore using atmospheric muon neutrinos, which to date have seen no significant sterile neutrino signal \cite{meows-prl,MEOWS, deepcore_sterile_2017}. Measurements of the electron neutrino spectrum at the MicroBooNE experiment, a liquid argon time projection chamber targeted by the same neutrino beam as MiniBooNE, also failed to reproduce an anomalous low-energy excess \cite{MicroBooNE:2021tya,MicroBooNE:2021pvo}. Other measurements from accelerator neutrino sources that are compatible with the absense of sterile neutrino mixing were performed by NO$\nu$A \cite{nova-sterile} and MINOS/MINOS+ \cite{minos:2019}. Several reactor neutrino experiments that use a near and far detector setup to cancel uncertainties of the initial neutrino flux also find no evidence for non-standard neutrino oscillations \cite{RENO:2020hva,STEREO:2019ztb,STEREO:2022nzk,DANSS:2018,DayaBay:2016}.  Global constraints on the unitarity of the PMNS matrix derived from non-anomalous results furthermore limit the magnitudes of    $|U_{e4}|^2$ and $|U_{\mu4}|^2$ to $\order{10^{-3}}$ and $\order{10^{-2}}$, respectively \cite{global_unitarity_Hu}. The amplitude $|U_{\tau4}|^2$ is currently only constrained to $\order{0.1}$. Furthermore, the number of relativistic neutrino species and the sum of neutrino masses  can  be constrained from cosmological observations. Sterile neutrinos could travel long distances unimpeded and therefore wash out the formation of structures at small scales in the early universe, which in turn would influence the power spectrum of the CMB and the formation of large structures.  Recent constraints from the Planck Collaboration for these parameters are $N_\mathrm{eff}=\num{2.99+-0.17}$ and $\sum m_\nu < \SI{0.1}{eV}$, and strongly disfavor the existence of sterile neutrinos within the standard $\Lambda$CDM paradigm \cite{Planck2018}.

The tension between highly significant anomalies in some experiments and strong exclusions from others is one of the most pressing problems in the field of neutrino physics. Its resolution necessitates the combination of independent and complementary measurements from various experiments probing different oscillation channels and energy ranges that are affected by different systematic uncertainties. This work uses atmospheric muon neutrinos to constrain the mixing amplitudes $|U_{\mu4}|^2$ and $|U_{\tau4}|^2$ under the assumption that $\Delta m^2_{41}\geq\SI{1}{eV^2}$. This is done by probing the $\nu_\mu\rightarrow\nu_\mu$ oscillation channel through an analysis of track-like events detected in IceCube DeepCore.
The dataset used for this analysis, described in \cite{verification-sample-prd}, incorporates numerous improvements over previous DeepCore studies in the event selection techniques and the modelling of systematic uncertainties, culminating in a greater statistical power and robustness than earlier results \cite{deepcore_sterile_2017}. 

\section{Atmospheric Muon Neutrino Sample}\label{sec:data-sample}

This study uses a dataset representing 7.5 years of livetime from the IceCube DeepCore detector. IceCube consists of \num{5160} downward-facing Digital Optical Modules (DOMs) that are deployed at depths between 1450-2450~m and distributed over 86 vertical cables\cite{icecube_detector_17}. The main array consists of 78 vertical cables that are arranged on a hexagonal lattice with a horizontal spacing of $\sim \SI{125}{m}$ between cables and a vertical spacing of \SI{17}{m} between DOMs. The remaining eight cables are located in the center of IceCube’s footprint with a tighter horizontal separation, between 40~m and 70~m. They contain DOMs with approximately 35\% higher quantum efficiency compared to the main array and are vertically spaced \SI{7}{m} apart. This central region of the detector forms the DeepCore sub-array with a fiducial volume of approximately \SI{10}{Mt} water equivalent. DeepCore is optimized for the observation of atmospheric neutrinos at energies $>\SI{5}{GeV}$ and uses the surrounding main array as a veto against muons originating from atmospheric showers that are the most significant background of this analysis \cite{DeepCore}. The data acquisition of DeepCore is triggered when a sufficient number of adjacent DOMs within the DeepCore volume record coincident signals that are within a 2.5~$\mu$s time window as described in \cite{deep-core-trigger}. 

The triggered events are passed through a series of cuts that reduce the background from random coincidences arising from detector noise and atmospheric muons while keeping most of the atmospheric neutrinos. This analysis uses the same event selection as a previous three-flavour atmospheric $\nu_{\mu}$ disappearance analysis described in \cite{verification-sample-prd}.

A first online filter at the South Pole vetoes events that are consistent with muons entering DeepCore from outside the detector based on hits recorded in the main IceCube array. Additional cuts of increasing complexity applied offline reduce the amount of background by approximately three orders of magnitude \cite{verification-sample-prd}. 

At this stage, the rate is approximately $\sim \SI{3}{\mu Hz}$ and the energy, zenith angle and flavour of each event is reconstructed. The zenith angle reconstruction is a simple geometric $\chi^2$-fit of the observed data to the expectation of Cherenkov light in the absence of light scattering as described in \cite{lowen-reco-paper}. The cosine of the zenith angle is used as a proxy for the distance traveled by a neutrino between its production in the atmosphere and its interaction in the detector, $L$. The energy reconstruction is based on a maximum likelihood reconstruction, where the expected charge for each DOM is taken from pre-computed tables \cite{Terliuk2018Measurement}. Both energy and zenith reconstructions are performed on each event once under a track-like event hypothesis, indicative of $\nu_{\mu}$ and $\bar{\nu}_{\mu}$ charged current (CC) interactions, and once under a cascade-like hypothesis, characteristic of $\nu_{e}$ CC, most $\nu_{\tau}$ CC and all neutral current (NC) interactions. For $\bar{\nu}_{\mu}$ CC interactions, which are the focus of this analysis, the energy reconstruction at a benchmark value of 20 GeV yields a bias of $0^{+5}_{-4}$~GeV. The zenith reconstruction at the same energy results in a bias of $6^{+12}_{-6}$ degrees. In both cases the bias is calculated as the mean of $X_\mathrm{reco} - X_\mathrm{true}$ and the range contains 50\% of the events around this mean. More details of the reconstruction performance for this sample can be found in \cite{verification-sample-prd}.

\begin{table*}[!t]
\caption{Observed and expected event rates for different types of particle interactions estimated at the best fit point of the analysis for the individual PID bins and in total.}\label{tab:event_rates}
\begin{tabular}{l|ccc|ccc|ccc}
                                                    & \multicolumn{3}{c|}{All PID}                              & \multicolumn{3}{c|}{Mixed PID}                            & \multicolumn{3}{c}{Track PID}                             \\ \hline
Event Type                                          & Events & Rate (1/10$^{6}$ s) & Fraction (\%) & Events & Rate (1/10$^{6}$ s) & Fraction (\%) & Events & Rate (1/10$^{6}$ s) & Fraction (\%) \\ \hline
$\nu_{\mu} + \overline{\nu}_{\mu}$ CC               & 17393  & 73.5                             & 79            & 6989   & 29.6                             & 65            & 10404  & 44.0                             & 93            \\
$\nu_{e} + \overline{\nu}_{e}$ CC                   & 1902   & 8.0                              & 8.6           & 1605   & 6.8                              & 15            & 298    & 1.3                              & 2.7           \\
$\nu_{\tau} + \overline{\nu}_{\tau}$ CC             & 599    & 2.5                              & 2.7           & 439    & 1.9                              & 4.1           & 160    & 0.7                              & 1.4           \\
$\nu_{\text{all}} + \overline{\nu}_{\text{all}}$ NC & 1128   & 4.8                              & 5.1           & 936    & 4.0                              & 8.7           & 192    & 0.8                              & 1.7           \\
Atm. Muons                                          & 971    & 4.1                              & 4.4           & 791    & 3.3                              & 7.3           & 180    & 0.8                              & 1.6           \\ \hline
All MC                                              & 21993  & 93.0                             &               & 10760  & 45.5                             &               & 11234  & 47.5                             &               \\
Data                                                & 21914  & 93.1                             &               & 10715  & 45.5                             &               & 11199  & 47.6                             &              
\end{tabular}
\end{table*}

The reduced $\chi^{2}$ from both track and cascade fits, together with the reconstructed track length and information about the location of the event in the detector, are passed into a Boosted Decision Tree (BDT) to calculate a particle identification (PID) score indicating how track-like, i.e. $\nu_{\mu}$ CC-like, an event signature appears. In this analysis, we refer to events with PID values between 0.75 and 1.0 as the ``tracks'' channel, which consists of $\sim93\%$ $\nu_\mu + \bar{\nu}_\mu$ CC events. Backgrounds in this channel consist mostly of atmospheric muons and $\nu_{\tau}$ CC events where the $\tau$-lepton decays to a muon with a branching ratio of $\sim17\%$. Events with a PID between 0.55 and 0.75 are referred to as the ``mixed'' channel, which still mostly consists of $\nu_\mu + \bar{\nu}_\mu$ CC events. Despite a lower purity of $\sim65\%$, this channel still enhances the sensitivity to sterile mixing due to the large number of $\nu_\mu + \bar{\nu}_\mu$ CC events, and is therefore used in the analysis. 
Since we do not consider cascade-like events with PID $< 0.55$ in this analysis, the number of electron neutrino interactions in the sample is reduced to below 10\%. This reduces the influence of the $\nu_{e}$ oscillation channels, such that $|U_{e4}|^2 = 0$ can be assumed without affecting the analysis. Finally, the events are binned in the reconstructed energy ($E_\mathrm{reco}$), the cosine of the reconstructed zenith angle ($\cos(\theta_z) \propto L$) and split by PID channel. We use the same binning as the three-flavour analysis in \cite{verification-sample-prd}. In this way, each bin is essentially an independent measurement in $L/E$ that can be used to probe atmospheric neutrino oscillations. 

With these criteria we select 21,914 well-reconstructed events for the analysis. The sample contains events spanning a reconstructed energy range from \SI{5}{GeV} to \SI{150}{GeV}, with a high purity of $\nu_\mu$ CC events and a small fraction of atmospheric muons. The total number of events observed in data and expected from MC in each PID channel is provided in \cref{tab:event_rates}. The rates expected from MC simulation are calculated at the best fit point of the analysis.

%%%%%%%%%% ANALYSIS

\section{Analysis Methodology}
This analysis employs a \emph{Monte-Carlo forward-folding} method to derive the expectation value of the event counts in the analysis histogram. A large set of MC simulated neutrino events corresponding to approximately $\SI{70}{yrs}$ of detector livetime has been generated and processed through the chain of filters previously described. These events are weighted according to the expected flux \cite{Honda:2015fha} multiplied by the oscillation probability and placed into a histogram with the same binning as the data. 

The weighting for each event can be adjusted using the neutrino oscillation parameters as well as a number of nuisance parameters corresponding to the systematic uncertainties in the atmospheric neutrino flux, the neutrino cross sections, the amount of atmospheric muon background and uncertainties of the detector properties. The parameter values are optimized with respect to a modified $\chi^2$ test statistic,
\begin{equation}\label{eq:mod_chi2}
\chi^2_{\mathrm{mod}} = \sum_{i \in \mathrm{bins}}^{}\frac{(N^{\mathrm{exp}}_i - N^{\mathrm{obs}}_i)^2}{N^{\mathrm{exp}}_i + (\sigma^{\mathrm{sim}}_i)^2} + \sum_{j \in \mathrm{syst}}^{}\frac{(s_j - \hat{s}_j)^2}{\sigma^2_{s_j}},
\end{equation}
that takes priors on the systematic uncertainties into account as well as the statistical uncertainty in the MC prediction, $\sigma_i^{\text{sim}}$. The index $i$ runs over every bin in the analysis histogram while $j$ runs over all nuisance parameters for which a Gaussian prior has been defined. The expected and observed counts in bin $i$ are $N_i^{\mathrm{exp}}$ and $N_i^\mathrm{obs}$. The variables $s_j$, $\hat{s}_j$ and $\sigma_{s_j}^2$ respectively denote the value of the systematic parameter $j$, its mean and its standard deviation. The total flux normalization is left unconstrained in this analysis, meaning that only effects on the shape of the signal are being considered.

\subsection{Calculation of Oscillation Probabilities}
\label{sec:sterile-neutrino-model}

We calculate neutrino oscillation probabilities using the \texttt{nuSQuIDS} \cite{squids, nusquids} package. This package computes state transition probabilities in the Interaction Picture, where the Hamiltonian is split into the time-independent vacuum oscillation part, $H_0$, and the variable interaction part, $H_1(t)$, such that 
\begin{equation}
    H(t) = H_0 + H_1(t)\;.
\end{equation}
In this picture, the probability to transition to state $i$ after the passage of time $t$,  $p_i(t)$, can be projected out of the state density matrix, $\bar{\rho}(t)$, with 
\begin{equation}
    p_i(t) = \text{Tr}(\bar{\Pi}^{(\alpha)}(t)\bar{\rho}(t))\,\label{eq:trace-operation}
\end{equation}
where $\bar{\Pi}^{(\alpha)}(t)$ is the projection operator for the flavour state $\alpha$. 
The state density at the time of detection is calculated by numerically integrating
\begin{equation}
    \partial_t \bar{\rho}(t)=-i[\bar{H}_1(t),\bar{\rho}(t)]\;.\label{eq:diff-eq}
\end{equation}
In both equations, the over-bar denotes the operator evolution
\begin{equation}
    \bar{O}(t)=e^{iH_0t}Oe^{-iH_0t}\;.\label{eq:operator-evolution}
\end{equation}
In the mass basis, the vacuum and interaction parts of the Hamiltonian in the 3+1 model can be written as
\begin{equation}
H_0 = \frac{1}{2E} \text{diag}(0, \Delta m^2_{21}, \Delta m^2_{31}, \Delta m^2_{41}) \label{eq:h0} 
\end{equation}
and
\begin{equation}
H_1 (t) = \frac{1}{\sqrt{2}} U_{3+1}^{\text{PMNS}, \dagger} \text{diag}( 2 V_\text{CC}(t), 0, 0, V_\text{NC}(t))U_{3+1}^\text{PMNS}\;,\label{eq:hi} 
\end{equation}
respectively.
The appearance of a non-zero neutral-current potential $V_\text{NC}$ in the interaction part of the Hamiltonian is a consequence of the fact that the sterile flavour state does not feel any potential. Therefore, the neutral-current interaction Hamiltonian is no longer diagonal and leads to non-trivial modifications of the neutrino propagation through matter. 
This method of calculating neutrino oscillation probabilities allows us to apply low-pass filters to the right hand side of \cref{eq:diff-eq} as well as the projection operator in \cref{eq:trace-operation} to greatly improve the efficiency of the calculation. In both cases, the filter replaces $\sin(\omega t)$ and $\cos(\omega t)$ terms that appear in the evaluation of the operator evolution \cref{eq:operator-evolution} by zero if the frequency $\omega = \Delta m^2_{i1}/(2E)$ lies above a given threshold. 

Oscillation lengths in the presence of an eV-scale mass splitting can be on the order of a few kilometers in the energy range relevant to this analysis. Therefore, we must take care of the assumed neutrino production height in the atmosphere. This is done by averaging the oscillation probability over a range of production heights by replacing the sine and cosine terms in \cref{eq:trace-operation} by their integral over the distance traveled by the neutrino. As a baseline we consider production heights between 10~km and 30~km of altitude. To minimize the effect of this approximation on the analysis, we do not consider events coming from more than $\approx6^\circ$ above the horizon. As a cross-check, we also varied the range of production height averaging to be between 1 and 20 km and found a negligible impact on the analysis. 

\begin{figure}
    \centering
    \includegraphics[width=1\linewidth]{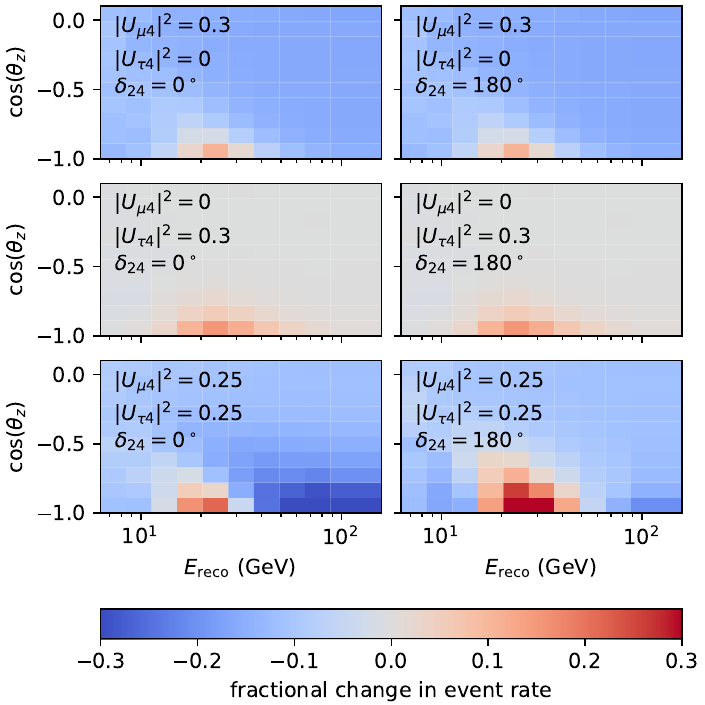}
    \caption{Sterile oscillation signal in the track channel of the analysis histogram for different combinations of $|U_{\mu 4}|^2$, $|U_{\tau 4}|^2$, and $\delta_{24}$. The fractional change is  $(N_{H_1} - N_{H_0}) / N_{H_0}$, where $N_{H_1}$ is the bin count for the sterile hypothesis and $N_{H_0}$ the bin count of the null hypothesis without sterile neutrinos.}
    \label{fig:signal-significance}
\end{figure}

In our chosen parametrization, 
\begin{align}
    U_{3+1}^\text{PMNS} =& R_{34}(\theta_{34}) \tilde{R}_{24}(\theta_{24}, \delta_{24}) \tilde{R}_{14}(\theta_{14}, \delta_{14}) R_{23}(\theta_{23}) \nonumber \\
    & \times \tilde{R}_{13}(\theta_{13}, \delta_{13}) R_{12}(\theta_{12})\;,
\end{align}
where $R_{ij}$ ($\tilde{R}_{ij}$) represents a (complex) rotation in the $ij$-plane.   The contribution from $\nu_{e}$ events is small enough that $|U_{e4}|^2$ can be neglected without impacting the constraints on the parameters of interest. Therefore, the parametrization of the mixing matrix simplifies to
\begin{align}
    |U_{\mu 4}|^2 &= \sin^2(\theta_{24}) \\
    |U_{\tau 4}|^2&= \sin^2(\theta_{34})\cos^2(\theta_{24})\;.
\end{align}
In contrast to an earlier DeepCore measurement \cite{Terliuk2018Measurement}, we marginalize over the sterile CP-violating phase $\delta_{24}$, and therefore test a more complete model, similar to \cite{ANTARES:2018rtf}.
\Cref{fig:signal-significance} shows the fractional change in the bin counts that would be produced in each bin of the track channel for different combinations of sterile oscillation parameters, where the null-hypothesis is the standard three-flavor oscillation scenario assuming NuFit~4.0 \cite{nufit40} global best fit parameters. In the energy range relevant for this analysis,  between 5 and 150~GeV, the observable effect of $|U_{\mu 4}|^2$ is an overall disappearance of muon neutrinos except for the region of maximum disappearance  between 15~GeV and 35~GeV as shown in the top row of \cref{fig:signal-significance}.  

The sensitivity of this analysis to $|U_{\tau4}|^2$ comes from the matter effects on neutrinos crossing the dense core of the Earth, where a non-zero value of  $|U_{\tau4}|^2$ leads to less disappearance of muon neutrinos between \SI{15}{GeV} and \SI{50}{GeV} as can be seen in the middle row in \cref{fig:signal-significance}. The ability to exploit this matter effect is a unique feature of atmospheric oscillation experiments and is what allows this measurement to be highly sensitive to $|U_{\tau4}|^2$ when compared to other types of neutrino oscillation experiments. The signal strength is greatest when both matrix elements $|U_{\mu 4}|$ and $|U_{\tau 4}|$ are non-zero, as correlations between them give rise to a signal that is more significant than a simple sum of the individual signals as can be seen in the bottom row of \cref{fig:signal-significance}. 

The energy resolution of the detector is not sufficient to resolve the rapid oscillation pattern that is produced by the heavy mass eigenstate at the assumed mass-splitting of \SI{1}{eV^2}. The signal shown in \cref{fig:signal-significance} is the result of these oscillation patterns being effectively averaged out in each bin. As a result of this averaging, the analysis is not sensitive to the precise value of the mass splitting between the sterile and active states and the constraints acquired from this measurement are valid for any mass splitting value $\geq \SI{1}{eV^2}$ up to approximately \SI{100}{eV^2}, at which point the heavy and active states begin to decohere \cite{atmo_decoherence}. This also simplifies the fit procedure because we can keep $\Delta m^2_{41}$ fixed to 1~eV$^2$.

\subsection{Systematic uncertainties}

The treatment of systematic uncertainties in this analysis follows a similar approach  as \cite{verification-sample-prd}. Here we provide an overview, highlighting the differences specific to this analysis. A summary of all systematic uncertainties and prior constraints, where applicable, is provided in \cref{tab:bfp-allpars}.

The baseline neutrino flux model \cite{Honda:2015fha} is adjusted to account for uncertainties in the primary cosmic ray spectral index \cite{Barr2006, nuflux_unc_manch}, as well as pion and kaon production uncertainties in air showers outlined in \cite{Barr2006} using the MCEq \cite{Fedynitch:2018cbl} package. Of the  subdivisions of the pion and kaon kinematic phase space described in \cite{Barr2006}, we include two parameters to account for pion uncertainties: $\Delta\pi^{\pm}$ [A-F] which modifies mostly low energy ($<10$ GeV) neutrino fluxes; and $\Delta\pi^{\pm}$ [I] to account for higher energy pion production uncertainties. A single parameter, $\Delta$K$^{+}$ [Y], is used to account for kaon production uncertainties. Variations in other parts of the kaon and pion phase space were found to be insignificant for this analysis.

Cross-section uncertainties for quasi-elastic and resonant neutrino scattering are parametrized based on variations of the respective axial masses in GENIE \cite{Andreopoulos:2015wxa}. To account for uncertainties in the modeling of deep inelastic scattering we follow the same method described in \cite{verification-sample-prd}, and include a parameter that interpolates between GENIE and CSMS \cite{csms-xsec} cross-sections. We also include an uncertainty of 20\% on the normalization of NC events to account for uncertainties in hadronization processes at the interaction vertex.

Similar to \cite{verification-sample-prd}, we estimate the baseline muon flux using the cosmic ray composition and flux from \cite{Gaisser:2011klf} and the Sibyll2.1 interaction model~\cite{Ahn_2009}. Given the minor contribution of atmospheric muons to the dataset used in this analysis, their uncertainty is accounted for by a simple scaling of the muon flux normalization which is left unconstrained in the fit. 

The largest contribution to the systematic error budget of this analysis comes from the uncertainties on the detector properties.
Just as in \cite{verification-sample-prd}, the systematic uncertainties related to detector calibration are parameterized by the optical efficiency of the DOMs, the average scattering and absorption coefficients of the natural glacial ice, and two parameters modeling the effects of the column of re-frozen ice surrounding the strings \cite{IceCube:2023ahv}. Prior constraints on these parameters, where applicable, are informed by calibration studies as described in \cite{verification-sample-prd}. 

Previously, these effects were quantified on the final histogram using linear regression through predictions from MC sets with varied detector parameters. The downside of this method is that the resulting linear functions are only valid at the flux and oscillation parameters that were chosen to calculate the histograms. For this analysis, we developed an entirely new approach to model detector effects. The new method uses the discrete MC sets to fit a classifier, which estimates the posterior probability that any given event belongs to a particular MC set, given true and reconstructed energy and zenith angles as well as the PID. These posterior probabilities can be used to re-weight each MC event according to its likelihood under a different realization of the detector properties. Because the relationship between true and reconstructed quantities is independent from the initial flux that produced the events, the resulting weight can therefore be applied under any flux and oscillation scenario without modification. The details of this new method are described in \cite{lowen-reco-paper}.

In total there are 18 nuisance parameters in the fit. Additional parameters, for example those related to the atmospheric neutrino flux and ice model uncertainties, were found to have a negligible impact on the analysis. The final set of nuisance parameters that impact this analysis is similar, though not identical, to the analysis presented in \cite{verification-sample-prd}. Although the two analyses use the same data sample and binning, there are several difference between them, such as the signal and the modelling of detector calibration uncertainties, which have resulted in slight differences between the nuisance parameters incorporated. However, importantly, the systematic uncertainties with the largest impact remain the same between both analyses. These are the detector calibration uncertainties, $\Delta\gamma_{\nu}$, and the atmospheric muon scale.

\section{Results}

\begin{table}[t]
    \caption{Best fit point of all free parameters of the analysis. The significance of the deviation from the nominal point (pull) is given for those parameters for which a Gaussian prior was defined. Parameters with Gaussian priors are allowed to vary within their $3\sigma$ range. If a uniform prior was applied to a parameter, its range is given in brackets instead. Blocks of phase space for pion and kaon yields denoted in brackets follow the definitions in \cite{Barr2006}.}
    \label{tab:bfp-allpars}
    \centering
    \begin{tabular}{lccc}
    \toprule
    Parameter & Best Fit Point & Prior & Pull ($\sigma$)\\
    \midrule
    \textbf{Detector}              &                          &   &  \\
    DOM eff. correction            &  $      108\%         $  &   $(100\pm10)\%$ &   0.812 \\
    Hole ice, rel. eff. $p_1$  &  $    0.0408          $  &   $[-0.15,0.1]$&   \\
    Hole ice, rel. eff. $p_0$  &  $    -0.589          $  &   $[-1.1,0.5]$&   \\
    Ice absorption                 &  $     98.8\%         $  &   $(100\pm5)\%$  &   -0.243 \\
    Ice scattering                 &  $     89.5\%         $  &   $(105\pm10)\%$ &   -1.546 \\
    \midrule 
    \textbf{Flux}                  &  \multicolumn{3}{c}{Changes w.r.t. Honda \emph{et al.}}  \\
$        \Delta \gamma_\nu$        &  $     0.091          $  &   0.0$\pm 0.1$      &  0.910 \\
   $\Delta\pi^{\pm}$ yields [A-F]  &  $     +10.6\%        $  &   0$\pm 63\%$     &  0.169 \\
   $\Delta\pi^{\pm}$ yields [I]    &  $     +44.6\%        $  &   0$\pm 61\%$     &  0.731 \\
   $\Delta K^{+}$ yields [Y]       &  $   -4.01\%          $  &   0$\pm 30\%$     &  -0.134 \\
   \midrule 
    \textbf{Cross-section}         &                          &   &  \\
$            M^\text{CCQE}_\text{A}$     &  $     -0.75\%           $  &   $0.99$~GeV~${}_{-15\%}^{+25\%}$    &  -0.050 \\
$           M^\text{CCRES}_\text{A}$     &  $    +1.9\%          $  &   1.12~GeV$ \pm$20\%     &  0.095 \\
$ \sigma_\text{NC}/ \sigma_\text{CC}  $          &  $         +0.005     $  &   1.0$\pm$0.2    &  0.024 \\
                        DIS CSMS   &  $     0.301          $  &   0.0$\pm 1.0$       &  0.301 \\
   \midrule
    \textbf{Oscillation} &                         &  & \\
$              \delta_{24}$        &  $       180    ^\circ$  &   $[0^\circ,180^\circ]$&\\
% $              \theta_{23}$  &  $      45.1    ^\circ$  &  \\
$      \sin^2(\theta_{23})$        &        0.502             &   $[0.12,0.88]$& \\
$\Delta m^{2}_{32}/\text{eV}^2$    &  $2.48\times 10^{-3}$    &   $[2,3]\times10^{-3}$&  \\
% $              \theta_{24}$        &  $      3.84    ^\circ$  &   uniform & \\
% $              \theta_{34}$        &  $      3.19    ^\circ$  &   uniform & \\
% $        \Delta m^{2}_{31}$  &  $      2.55\times 10^{-3}\,\text{eV}^2$  &  \\
$|U_{\mu 4}|^2$ & 0.0045 & $[0.0,0.72]$&  \\
$|U_{\tau 4}|^2$ & 0.0031 & $[0.0,0.72]$&  \\
   \midrule
    \textbf{Atm. muons}            &                          &  & \\
Atm. $\mu$ scale                   &  $      1.9         $    &  $[0.0,3.0]$& \\
   \midrule
    \textbf{Normalization}         &                          &  & \\
$A_{\mathrm{eff.}}$ scale                   &  $     0.74         $    &  $[0.2,2.0]$& \\
    \bottomrule
    \end{tabular}
\end{table}

The result of the measurement is compatible with the absence of sterile neutrino mixing and the marginalized constraints for the matrix elements at the 90\% and 99\% confidence levels are
\begin{equation}
\begin{aligned}
    |U_{\mu 4}|^2 & < 0.0534\;(90\%\;\mathrm{CL}),\;0.0752\;(99\%\;\mathrm{CL})\;,\\
    |U_{\tau 4}|^2 & < 0.0574\;(90\%\;\mathrm{CL}),\;0.0818\;(99\%\;\mathrm{CL})\;.\\
\end{aligned}
\end{equation}

\begin{figure}[t]
    \centering
    \includegraphics[width=1\linewidth]{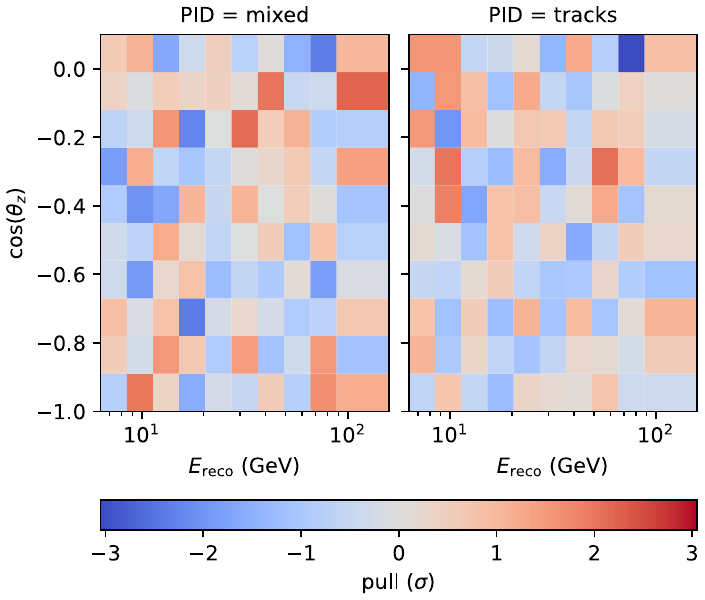}
    \caption{Bin-wise significance of the deviations between the observed data and the MC prediction at the best fit point of the analysis. The values shown include the Poisson error of the data as well as the error due to finite MC statistics.}
    \label{fig:binwise-pulls-bfp}
\end{figure}

\begin{figure}[t]
    \centering
    \includegraphics[width=1\linewidth]{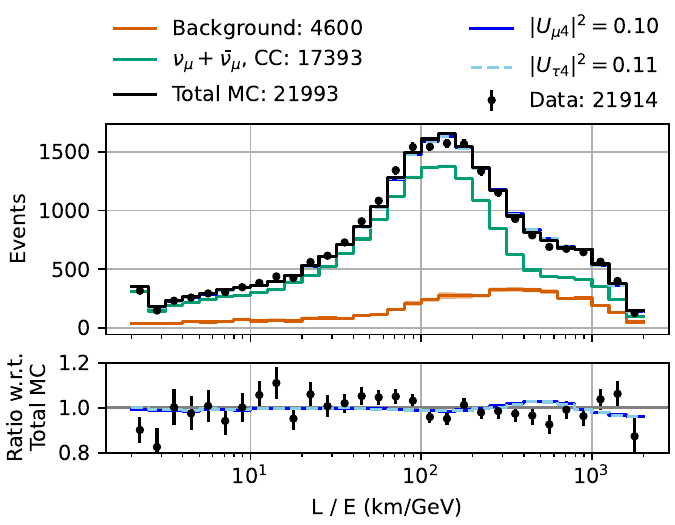}
    \caption{Post-fit distribution of $L/E$, compared to the observed data. The background is the sum of atmospheric muon events and all neutrino events except for charged-current $\nu_\mu$ interactions. The legend shows the number of events in each histogram. Errors include Poisson errors from data as well as the uncertainties due to MC statistics. The outermost bins include overflow events. Alternative hypotheses for $|U_{\mu 4}|^2$ and $|U_{\tau 4}|^2$ are shown in blue after marginalizing over all nuisance parameters.}
    \label{fig:post-fit-LovE}
\end{figure}

In \cref{fig:binwise-pulls-bfp}, we show the significance of the deviation between the observed data and the MC prediction at the best fit point of the analysis. Overall we observe good agreement between data and MC, with a p-value of 22.5\%. This is also demonstrated by \cref{fig:post-fit-LovE}, which shows the data from each bin projected in $L/E$. For reference we show two additional models with $|U_{\mu 4}|^2$ and $|U_{\tau 4}|^2$ individually set to values at approximately 99\% CL. The strong degeneracy between these two parameters when one is fixed to zero is demonstrated by the large overlap between the models. As discussed in Section~\ref{sec:sterile-neutrino-model}, the fast oscillations from a 1~eV$^{2}$ scale additional mass splitting are averaged out in this $L/E$ regime, and the signal is instead an overall distortion of the spectrum, particularly for long baselines as previously shown in \cref{fig:signal-significance}.

\begin{figure}[t]
    \centering
    \includegraphics[width=\linewidth]{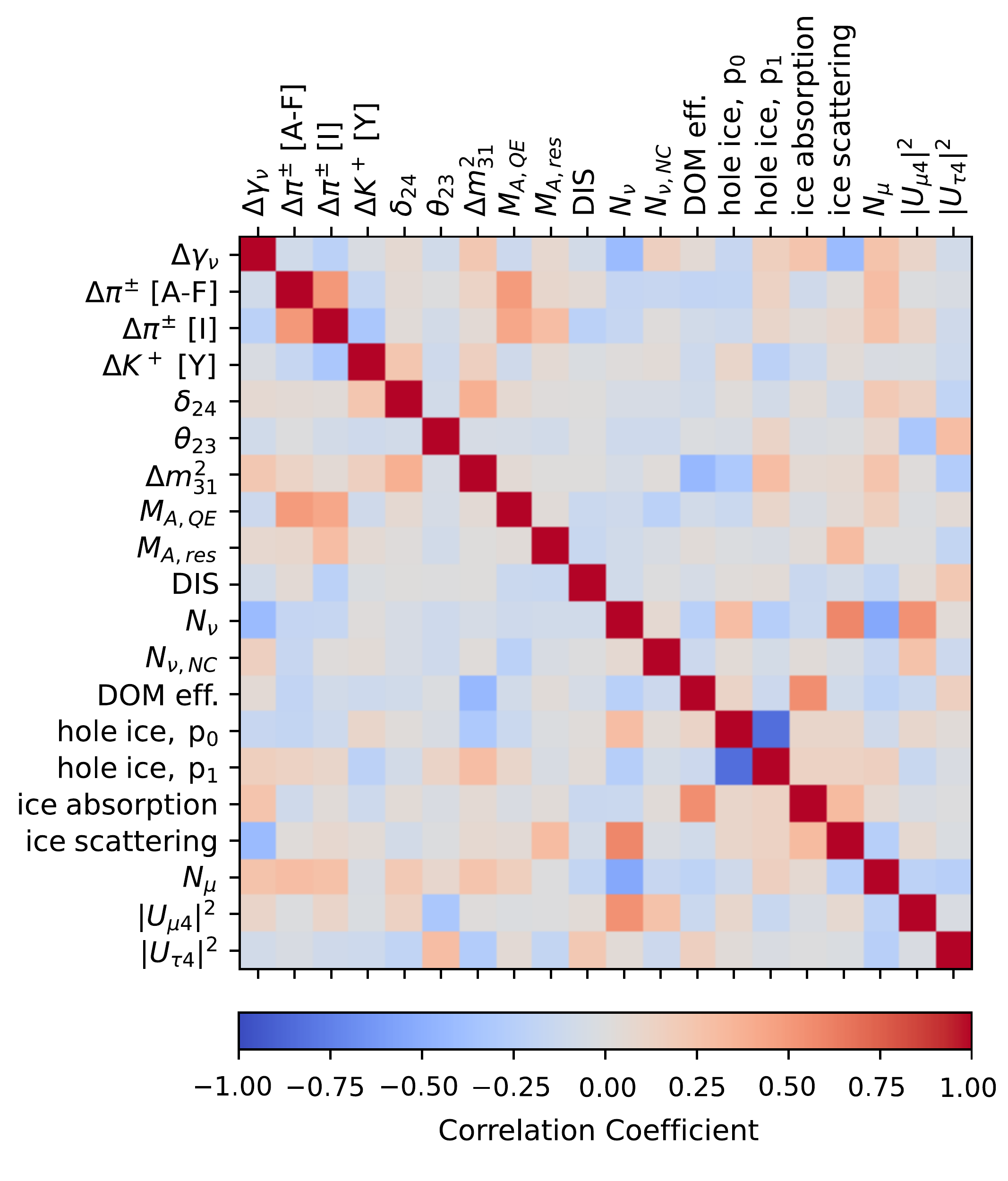}
    \caption{Pearson correlation coefficients between all free parameters of the analysis, calculated at the best fit point of the analysis.}
    \label{fig:parameter-correlations}
\end{figure}

The best fit points of all nuisance parameters, shown in \cref{tab:bfp-allpars}, are within prior expectations. The values of the atmospheric neutrino oscillation parameters $\theta_{23}$ and $\Delta m^2_{32}$, which are treated as free nuisance parameters in this analysis, fit within the $1\sigma$ range of the result in \cite{verification-sample-prd}, although our model fits slightly closer to maximal three-flavor mixing. This is likely due to the slight under-fluctuation of data observed in \cref{fig:post-fit-LovE} for $10^2\leq L/E \leq 10^3$. The atmospheric muon scale fits to a value of 1.9, almost doubling the rate. This is likely due to a combination of statistical fluctuations, given the small number of muons in the sample, and an under-estimation of the baseline atmospheric muon flux, which has been observed in other measurements \cite{Riehn:2024prp,KM3NeT:2024buf}.
Our best fit neutrino normalization is also lower than in \cite{verification-sample-prd}, which can be explained by changes in several correlated parameters, which are shown in \cref{fig:parameter-correlations}. In particular, the neutrino normalization is negatively correlated with the atmospheric muon flux and with the spectral index of the neutrino flux and positively correlated with the scattering coefficient of the ice. In each of these parameters, our fit results have changed with respect to \cite{verification-sample-prd} in a way that compensates for the lower normalization.

Since $\delta_{24}$ is treated as a free parameter in the fit, these results are valid for both the normal and inverted neutrino mass orderings due to the approximate degeneracy between the mass ordering and the sign of $\cos(\delta_{24})$ as described in \cite{ANTARES:2018rtf}. Compared to the previous DeepCore analysis, the limits at 90\% CL are improved by a factor of 2.1 and 2.6 for $|U_{\mu 4}|^2$ and $|U_{\tau 4}|^2$, respectively. The limit on $|U_{\tau 4}|^2$ in particular is competitive with limits obtained from global unitarity constraints of the PMNS matrix \cite{global_unitarity_Hu}. The improved sensitivity is largely due to the increase in statistics with a larger data sample, with additional improvements derived from improved detector calibration and treatment of systematic uncertainties. 

\begin{figure}[t!]
    % adding a little bit of vertical space so that the figure doesn't hug the text
    \vspace{0.1cm}
    \includegraphics{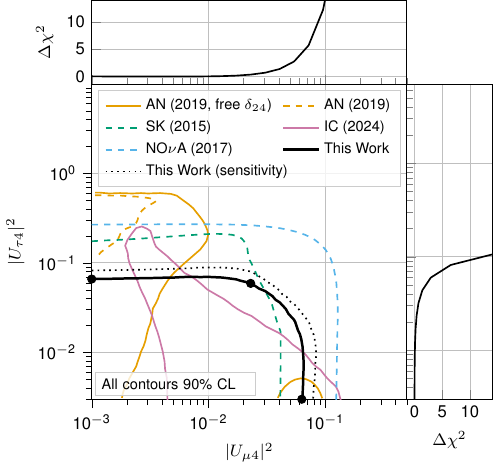}
    \caption{Contour of the 90\% CL limit of this analysis compared to measurements from the ANTARES \cite{ANTARES:2018rtf}, Super-Kamiokande \cite{Super-Kamiokande:2014ndf} and NO$\nu$A \cite{nova-sterile} experiments and a recent high-energy IceCube oscillation study \cite{placeholder-meows-th34}. Black dots along the contour of this work indicate where coverage spot-checks were run. The dotted line shows the expected sensitivity of this analysis. Results shown as dashed lines assume $\delta_{24}=0$.}
    \label{fig:result-contours}
\end{figure}

 We performed a scan over $|U_{\mu 4}|^2$ and $|U_{\tau 4}|^2$ with respect to the $\Delta \chi^2 = \chi^2_\mathrm{mod,\;best\;fit}-\chi^2_\mathrm{mod,\;scan\;point}$ test statistic and estimated the 90\% CL contours using Wilks' theorem assuming two degrees of freedom. The results are shown in \cref{fig:result-contours}. We ran spot-checks of the coverage of the test statistic distribution using 200 pseudo-data trials of randomly fluctuated histograms on three points along the contour as shown in \cref{fig:result-contours}. We found that the 90\% quantile of the empirical test statistic distribution was lower than the value given by Wilks' theorem on all test points. Thus, the contours drawn in \cref{fig:result-contours} are a conservative estimate of the correct limits. The limits obtained from the observed data are more stringent than the expected sensitivity, which is also shown in \cref{fig:result-contours}. This is due to the under-fluctuation of observed events in the oscillation region that was also reported in \cite{verification-sample-prd} and deemed to be compatible with statistical fluctuations therein. Since any non-zero sterile mixing amplitude leads to an increase in the bin counts in the energy range of maximal muon neutrino disappearance as shown in \cref{fig:signal-significance}, a statistical under-fluctuation in these bins causes a stronger preference for the null hypothesis and therefore explains the more stringent limits observed in this work. 

Compared to other recent 3+1 sterile neutrino searches by IceCube \cite{placeholder-meows-th34,meows-bdt}, which leverage the MSW resonance effect at TeV energies, our methodology investigates sterile neutrino mixing in the 5 - 150 GeV energy range, which is far from this resonance condition.  In addition to the physical effect being probed, systematic uncertainties, especially in neutrino cross-sections, vary markedly with energy. The allowed region of phase space from the most directly comparable high-energy IceCube search is shown in \cref{fig:result-contours}. Pursuing both analysis methodologies exploits the full energy range that is observable with the IceCube DeepCore detector, and provides complementary approaches to investigate the 3+1 sterile neutrino landscape. 

In summary, the measurement described in this Letter adds a new non-observation to the global picture of 3+1 fits with new competitive limits on $|U_{\mu4}|^2$ and $|U_{\tau4}|^2$.  This result adds critical information to the ongoing discourse in neutrino physics by addressing the tension between experimental anomalies suggesting active-sterile neutrino mixing and the strong exclusions from other measurements. The sensitivity of this study to $|U_{\tau4}|^2$, leveraging matter effects in atmospheric neutrino oscillations, exemplifies the importance of diverse experimental approaches in resolving the complex puzzle of neutrino behavior.

%TC:ignore
\begin{acknowledgments}

The authors gratefully acknowledge the support from the following agencies and institutions:
USA {\textendash} U.S. National Science Foundation-Office of Polar Programs,
U.S. National Science Foundation-Physics Division,
U.S. National Science Foundation-EPSCoR,
U.S. National Science Foundation-Office of Advanced Cyberinfrastructure,
Wisconsin Alumni Research Foundation,
Center for High Throughput Computing (CHTC) at the University of Wisconsin{\textendash}Madison,
Open Science Grid (OSG),
Partnership to Advance Throughput Computing (PATh),
Advanced Cyberinfrastructure Coordination Ecosystem: Services {\&} Support (ACCESS),
Frontera computing project at the Texas Advanced Computing Center,
U.S. Department of Energy-National Energy Research Scientific Computing Center,
Particle astrophysics research computing center at the University of Maryland,
Institute for Cyber-Enabled Research at Michigan State University,
Astroparticle physics computational facility at Marquette University,
NVIDIA Corporation,
and Google Cloud Platform;
Belgium {\textendash} Funds for Scientific Research (FRS-FNRS and FWO),
FWO Odysseus and Big Science programmes,
and Belgian Federal Science Policy Office (Belspo);
Germany {\textendash} Bundesministerium f{\"u}r Bildung und Forschung (BMBF),
Deutsche Forschungsgemeinschaft (DFG),
Helmholtz Alliance for Astroparticle Physics (HAP),
Initiative and Networking Fund of the Helmholtz Association,
Deutsches Elektronen Synchrotron (DESY),
and High Performance Computing cluster of the RWTH Aachen;
Sweden {\textendash} Swedish Research Council,
Swedish Polar Research Secretariat,
Swedish National Infrastructure for Computing (SNIC),
and Knut and Alice Wallenberg Foundation;
European Union {\textendash} EGI Advanced Computing for research;
Australia {\textendash} Australian Research Council;
Canada {\textendash} Natural Sciences and Engineering Research Council of Canada,
Calcul Qu{\'e}bec, Compute Ontario, Canada Foundation for Innovation, WestGrid, and Digital Research Alliance of Canada;
Denmark {\textendash} Villum Fonden, Carlsberg Foundation, and European Commission;
New Zealand {\textendash} Marsden Fund;
Japan {\textendash} Japan Society for Promotion of Science (JSPS)
and Institute for Global Prominent Research (IGPR) of Chiba University;
Korea {\textendash} National Research Foundation of Korea (NRF);
Switzerland {\textendash} Swiss National Science Foundation (SNSF).

\end{acknowledgments}

\bibliographystyle{apsrev4-2}
\bibliography{refs}
%TC:endignore
\end{document}